\documentclass[a4paper]{article} 

\usepackage[utf8]{inputenc}
\usepackage[english]{babel}
\usepackage{pifont}
\usepackage{amsmath,amsbsy,bm,amssymb,mathtools,dsfont,etoolbox,braket}
\usepackage{graphicx,float,subcaption}
\usepackage{multirow,makecell,array,tabularx}
\usepackage[dvipsnames]{xcolor}
\usepackage{tikz,pgfplots,pgfplotstable}
\usepackage[mode=buildnew]{standalone} 
\usepackage{diagbox}
\usepackage{siunitx}
\usepackage{xspace}
\usepackage{wrapfig,booktabs}
\usepackage[numbers]{natbib}
\usepackage{textgreek}

\makeatletter\@ifclassloaded{svjour3}{\input{main-svjour3}}{
\usepackage[font=small,labelfont=bf]{caption}
\usepackage[top=3cm,bottom=3cm,left=3cm,right=3cm]{geometry}
\usepackage{orcidlink}
\usepackage{hyperref}
\usepackage{cleveref}

\newcommand{\keywords}[1]{}
\newenvironment{acknowledgements}{\phantomsection\addcontentsline{toc}{section}{Acknowledgements}\section*{Acknowledgements}}{}
\newcommand{\reftoc}{\phantomsection\addcontentsline{toc}{section}{\refname}}

\author{Raoul~Heese\textsuperscript{1*}\orcidlink{0000-0001-7479-3339}, Moritz~Wolter\textsuperscript{2}, Sascha~Mücke\textsuperscript{3}\orcidlink{0000-0001-8332-6169},\\Lukas~Franken\textsuperscript{4}, Nico~Piatkowski\textsuperscript{4}\orcidlink{0000-0002-6334-8042}}
\date{%
	\small%
	\textsuperscript{1}\itshape Fraunhofer Center for Machine Learning and Fraunhofer Institute for Industrial Mathematics ITWM \upshape\\%
	\textsuperscript{2}\itshape Fraunhofer Center for Machine Learning and Fraunhofer Institute for Algorithms and Scientific Computing SCAI \upshape\\%
	\textsuperscript{3}\itshape Artificial Intelligence Group, TU Dortmund University \upshape\\%
	\textsuperscript{4}\itshape Fraunhofer Institute for Intelligent Analysis and Information Systems IAIS \upshape\\%
	\textsuperscript{*}\itshape raoul.heese@itwm.fraunhofer.de \upshape\\[2ex]%
	\normalsize%
	\today%
}
}\makeatother 

\setcitestyle{authoryear,open={(},close={)}}

\hypersetup{hidelinks}

\pgfplotsset{compat=newest}
\usetikzlibrary{calc,math,decorations.pathreplacing,decorations.pathmorphing}
\usepgfplotslibrary{groupplots,fillbetween}

\crefname{figure}{Fig.}{Figs.}
\Crefname{figure}{Fig.}{Figs.}
\crefname{table}{Tab.}{Tabs.}
\Crefname{table}{Tab.}{Tabs.}
\crefname{equation}{Eq.}{Eqs.}
\Crefname{equation}{Eq.}{Eqs.}

\newcommand{\ie}{i.\,e.\xspace}
\newcommand{\eg}{e.\,g.\xspace}
\newcommand{\cf}{cf.\xspace}
\newcommand{\iid}{i.i.d.\xspace}
\newcommand{\cellrot}[1]{\begin{tabular}{@{}c@{}}#1\end{tabular}}
\newcommand{\accept}{\ding{51}}
\newcommand{\reject}{\ding{55}}
\newcommand{\passed}{passed}

\title{On the effects of biased quantum random numbers on the initialization of artificial neural networks} 

\begin{document}
		
\maketitle

\begin{abstract}
Recent advances in practical quantum computing have led to a variety of cloud-based quantum computing platforms that allow researchers to evaluate their algorithms on noisy intermediate-scale quantum (NISQ) devices. A common property of quantum computers is that they can exhibit instances of true randomness as opposed to pseudo-randomness obtained from classical systems. Investigating the effects of such true quantum randomness in the context of machine learning is appealing, and recent results vaguely suggest that benefits can indeed be achieved from the use of quantum random numbers. To shed some more light on this topic, we empirically study the effects of hardware-biased quantum random numbers on the initialization of artificial neural network weights in numerical experiments. We find no statistically significant difference in comparison with unbiased quantum random numbers as well as biased and unbiased random numbers from a classical pseudo-random number generator. The quantum random numbers for our experiments are obtained from real quantum hardware.
\keywords{Quantum Computing, Random Number Generation, Neural Networks, Machine Learning}
\end{abstract}

\section{Introduction} \label{sec: introduction}
The intrinsic non-deterministic nature of quantum mechanics \citep{kofler2010} makes random number generation a native application of quantum computers. It has been exemplarily studied in \citet{bird2020} how such quantum random numbers can affect stochastic machine learning algorithms. For this purpose, electron-based superposition states have been prepared and measured on quantum hardware to create random 32-bit integers. These numbers have subsequently been used to initialize the weights in neural network models and to determine random splits in decision trees and random forests. \citeauthor{bird2020} have observed that quantum random numbers can lead to superior results for certain numerical experiments in comparison with classically\footnote{We use the term ``classical'' in the sense of the physics community to distinguish deterministically behaving entities from the realm of classical physics from those governed by the non-deterministic rules of quantum physics \citep{norsen2017}.} generated pseudo-random numbers.\par
However, the authors have not further explained this behavior. In particular, they have not discussed the statistical properties of the generated quantum numbers. Due to technical imperfections and physical phenomena like decoherence and dissipation, measurement results from a quantum computer might in fact significantly deviate from idealized theoretical predictions \citep{tamura2020,shikano2020,tamura2021}. This raises the question of whether it is not the superiority of the quantum random number generator to sample perfectly random from the uniform distribution that leads to the observed effect, but instead its ability to sample bit strings from a very particular distribution that is imposed by the quantum hardware.\par
We therefore revisit this topic in the present manuscript and generate biased random numbers using real quantum hardware, where the specifics of the bias are determined by the natural imperfections of the hardware itself. The bias is therefore not under our control and even beyond our full understanding. With this approach, we aim to better comprehend the effects observed by \citeauthor{bird2020} for an analogous setup and explore the resulting implications. Summarized, our main goal is to further study the results of that work and to analyze the effects of quantum and classical random numbers with and without biases on neural network initialization. Our analysis is mainly based on numerical experiments and statistical tests.\par
The structure of the remaining paper is as follows. In \cref{sec:background}, we briefly summarize the background of the main ingredients of our work, namely quantum computing and random number generation. Subsequently, we present the setup of our quantum random number generator and discuss the statistics of its results in \cref{sec:biased qrng}. In \cref{sec:experiments}, we study the effects of the generated quantum random numbers on artificial neural network weight initialization using numerical experiments. Finally, we close with a conclusion.

\section{Background} \label{sec:background}
In the following, we provide a brief introduction to quantum computing and random number generation without claiming to be exhaustive. For more in-depth explanations, we refer to the cited literature.

\subsection{Quantum computing}
Quantum mechanics is a physical theory that describes objects at the scale of atoms and subatomic particles, \eg, electrons and photons \citep{norsen2017}. An important interdisciplinary subfield is quantum information science, which considers the interplay of information science with quantum effects and includes the research direction of quantum computing \citep{nielsen2011}. 

\subsubsection{Quantum devices}
A quantum computer is a processor which utilizes quantum mechanical phenomena to process information \citep{benioff1980,nas2019}. Theoretical studies show that quantum computers are able to solve certain computational problems significantly faster than classical computers, for example, in the fields of cryptography \citep{pirandola2020} and quantum simulations \citep{georgescu2014}. Recently, different hardware solutions for quantum computers have been realized and are steadily improved. For example, superconducting devices \citep{huang2020} and ion traps \citep{bruzewicz2019} have been successfully used to perform quantum computations. However, various technical challenges are still unresolved so that the current state of technology, which is subject to substantial limitations, is also phrased as \emph{noisy intermediate-scale quantum} (NISQ) computing \citep{preskill2018}. Nevertheless, quantum supremacy on NISQ devices has already been verified experimentally for a specialized task of randomized sampling \citep{boixo2018,wu2021}.\par
There are different theoretical models to describe quantum computers, typically used for specific hardware or in different contexts. We only consider the quantum circuit model, in which a computation is considered as a sequence of \emph{quantum gates} and the quantum computer can consequently be seen as a \emph{quantum circuit} \citep{nielsen2011}. In contrast to a classical computer, which operates on electronic bits with a well-defined binary state of either 0 or 1, a quantum circuit works with \emph{qubits}. A qubit is described by a quantum mechanical state, which can represent a binary 0 or 1 in analogy to a classical bit. In addition, however, it can also represent any \emph{superposition} of these two values. Such a quantum superposition is a fundamental principle of quantum mechanics and cannot be explained with classical physical models. Moreover, two or more qubits can be \emph{entangled} with each other. Entanglement is also a fundamental principle of quantum mechanics and leads to non-classical correlations \citep{bell2004}.\par
In order to illustrate the aforementioned fundamental quantum principles and to connect them with well-known notions from the field of machine learning, one can consider the following intuitive (but physically inaccurate) simplifications: Superposition states can be understood as probability distributions over a finite state space, while entanglement amounts to high-order dependencies between univariate random variables. This intuition particularly emphasizes the close relationship between quantum mechanics and probability theory.\par
\begin{figure}[!t]
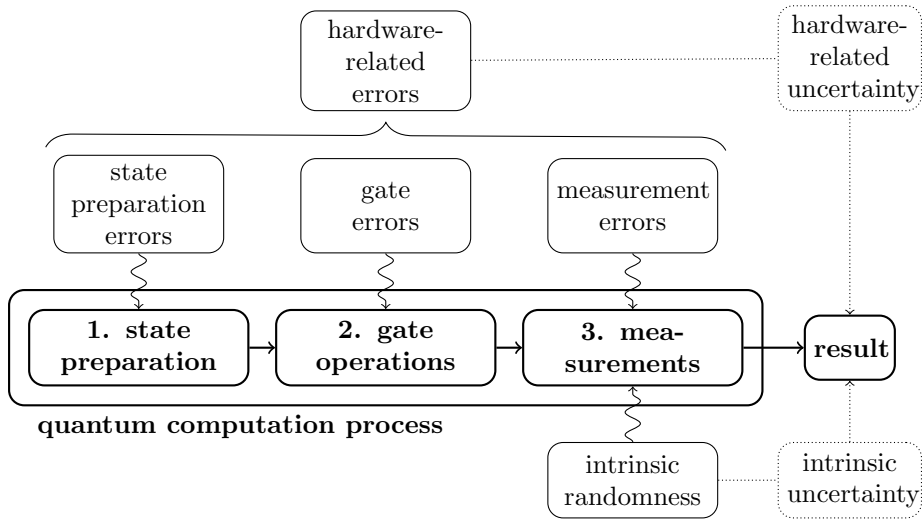

\centering
\includestandalone{./fig_qcp}
\caption{Sketch of the three-step quantum computation process consisting of an initial state preparation, a sequence of gate operations and a final measurement, which yields the result of the computation. Also shown are the errors associated with each step in the computation process: the state preparation errors, the gate errors, and the measurement errors, respectively. They are all hardware-related errors, which can in principle be reduced (or even eliminated) by technological advances. These errors can cause a hardware-related uncertainty (statistical and systematic) of the computation result. On the other hand, the intrinsic randomness of quantum mechanics emerging at the time of the measurement causes an intrinsic uncertainty of the computation result, which is an integral part of quantum computing and can be exploited to construct QRNGs.}
\label{fig:qcp}
\end{figure}

Any quantum computation can be considered as a three-step process, which is sketched in \cref{fig:qcp}. First, an initial quantum state of the qubits is prepared, usually a low-energy ground state. Second, a sequence of quantum gates deterministically transforms the initial state into a final quantum state. Third, a measurement is performed on the qubits to determine an outcome. When a qubit is measured, the result of the measurement is always either 0 or 1, but the observation is non-deterministic with a probability depending on the quantum state of the qubit at the time of the measurement.\par
In this sense, a quantum computation includes an intrinsic element of randomness. This randomness is in particular not a consequence of lack of knowledge about the quantum system, but an integral part of quantum mechanics itself. In constrast to classical mechanics, where complete knowledge about the intitial state of a system allows to infer all later (and earlier) states, complete knowledge about a quantum mechanical state does not generally allow the prediction of a single measurement outcome, but only its probability as determined by Born's rule \citep{norsen2017}. The non-deterministic nature of quantum mechanics relies on the assumption that there are no so-called \emph{hidden variables} whose knowledge would lead to a deterministic behavior \citep{norsen2017}. Various theoretic and experimental evidences, for example based on Bell's theorem \citep{bell2004} or the Kochen–Specker theorem \citep{kochen1975}, strongly suggest that there are no such hidden variables. However, a conclusive answer to the question of quantum non-determinism is still in scientific discourse. For a more detailed discussion about this topic, we refer to \citet{bera2017} and references therein. Since our work concerns the practical application of random numbers in machine learning algorithms and a theoretical provability of their randomness from first principles is beyond the scope of this paper, we presume in the following that quantum mechanics is indeed intrinsically non-deterministic for all purposes considered.\par
NISQ devices, as their name suggests, are typically only capable of computing noisy results. A fundamental reason is that the quantum computer, despite all technical efforts, is not perfectly isolated and interacts (weakly) with its environment. In particular, there are two major effects of the environment that can contribute to computational errors, namely dissipation and decoherence in the sense of dephasing \citep{zurek2007,vacchini2016}. Dissipation describes the decay of qubit states of higher energy due to an energy exchange with the environment. Decoherence, on the other hand, represents a loss of quantum superpositions as a consequence of environmental interactions. Typically, decoherence is more dominating than dissipation. Beyond these typical effects, other (possibly unknown) influences can occur, which can lead to additional uncertainties.\par
To compensate the resulting computational errors to a certain extend, error correction can be used \citep{roffe2019}. However, it is generally not possible to completely eliminate statistical (also called \emph{aleatoric}) or systematic (also called \emph{epistemic}) uncertainties, which might originate from quantum and classical effects, respectively. Therefore, quantum algorithms must be designed sufficiently robust for practical applications on NISQ hardware.\par 
In \cref{fig:qcp}, we briefly outline different error sources in the quantum computation process. Specifically, each computation step is affected by certain hardware-related errors, which are referred to as state preparation errors, gate errors, and measurement errors, respectively \citep{nachman2021}. All of them are a consequence of the imperfect physical hardware and they are non-negligible for NISQ devices \citep{leymann2020}. The resulting hardware-related uncertainty might be both statistical and systematic. In addition, the final measurement step is also affected by the intrinsic randomness of quantum mechanics. The measurement ultimately yields a computation result that contains two layers of uncertainty \citep{heese2014}: First, the uncertainty caused by the hardware-related errors, and second, the uncertainty caused by the intrinsic randomness. While technological advances (like better hardware and improved algorithm design) can in principle reduce (or even eliminate) hardware-related errors and thus the hardware-related uncertainty, the intrinsic uncertainty is an integral part of quantum computing. It is this intrinsic uncertainty which can be exploited to construct QRNGs.

\subsubsection{Quantum machine learning}
In a machine learning context, we may identify a quantum circuit with a parameterizable probability distribution over all possible measurement outcomes, where each measurement of the circuit draws a sample from this distribution. The interface between quantum mechanics and machine learning can be attributed to the field of \emph{quantum machine learning} \citep{biamonte2017}. A typical use case is the processing of classical data using algorithms that are fully or partially computed with quantum circuits, which is also called \emph{quantum-enhanced machine learning} \citep{dunjko2016}.\par
The noisy nature of NISQ devices presents a challenge for machine learning applications. On the other hand, the probabilistic nature of quantum computing can be related to the statistical background of machine learning algorithms, for which the understanding and modeling of uncertainty is crucial. A review about different types of uncertainty in machine learning and how to typically deal with them can for example be found in \citet{huellermeier2021}.

\subsection{Random number generation}
For many machine learning methods, random numbers are a crucial ingredient and therefore random number generators (RNGs) are an important tool. Examples include sampling from generative models like generative adversarial networks, variational autoencoders or Markov random fields, parameter estimation via stochastic optimization methods, as well as randomized regularization and validation techniques, randomly splitting for cross-validation, drawing of random mini-batches, and computing a stochstic gradient, to name a few. Randomness also plays an important role in non-deterministic optimization algorithms or the initialization of (trainable) neural network parameters \citep{glorot2010,he2015}.\par
At its core, a RNG performs random coin tosses in the sense that it samples from a uniform distribution over a binary state space (or, more generally, a discrete state space of arbitrary size). Given a sequence of randomly generated bits, corresponding integer or floating-point values can be constructed straightforwardly.

\subsubsection{Classical RNGs}
In the classical world, there are two main types of random number generators. Pseudo-random number generators (PRNGs) represent a class of algorithms to generate a sequence of apparently random (but in fact deterministic) numbers from a given \emph{seed} \citep{james2020}. In other words, the seed fully determines the order of the bits in the generated sequence, but the statistical properties of the sequence (\eg, mean and variance) are independent of the seed (as determined by the underlying algorithm). We remark that PRNGs can also be constructed based on machine learning algorithms \citep{pasqualini2020}.\par
The more advanced true random number generators (TRNGs) are hardware devices that receive a signal from a complex physical process, which is unpredictable for all practical purposes, to extract random numbers \citep{yu2019}. A multitude of physical effects can be used as sources of entropy for TRNGs, with only some of them directly linked to quantum phenomena. For example, metastability in latches can be exploited in specialized electrical circuits (CMOS devices) to yield random bits \citep{tokunaga2008,holleman2008}. Usually, such setups are built to calibrate themselves to account for hardware-inherent bias effects. Multiple of these self-calibrating entropy sources can be combined to further increase the cryptographic quality \citep{mathew2015}. Other approaches make use of ring oscillators to source randomness from timing jitter \citep{kim2017}, or exploit random telegraph noise to produce bit streams \citep{puglisi2018,brown2020}.\par
For TRNGs, the lack of knowledge about the observed physical system induces randomness, but it cannot be guaranteed in principle that the dynamics of the underlying physical system are unpredictable (if quantum effects are not sufficiently involved). Likewise, the statistical properties of the generated random sequence are not in principle guaranteed to be constant over time since they are subject to the hidden process.\par
Independent of their source, random numbers have to fulfill two properties: First, they have to be truly random (\ie, the next random bit in the sequence must not be predictable from the previous bits) and second, they have to be unbiased (\ie, the statistics of the random bit sequence must correspond to the statistics of the underlying uniform distribution). In other words, they have to be secure and reliable. A ``good'' RNG has to produce numbers that fulfill both requirements. In practice, it is difficult to rigorously proof the quality of RNGs. For a bit sequence of finite length, there is no formal method to decide its randomness with certainty. On the other hand, an infinite bit sequence cannot be tested in finite time \citep{khrennikov2015}. Therefore, statistical test are typically used to check specific properties of RNGs with a certain confidence.\par
Typically, statistical tests are organized in the form of test suites (\eg, the \emph{NIST Statistical Test Suite} described in \citealp{rukhin2010}) to provide a comprehensive statistical screening. A predictive analysis based on machine learning methods can also be used for a quality assessment \citep{cai2020}. It remains a challenge to certify classical RNGs in terms of the aforementioned criteria \citep{balasch2018} to, \eg, ensure cryptographical security.\par
When implementing learning and related algorithms, PRNGs are typically used. Despite the broad application of randomness in machine learning, the apparent lack of research regarding the particular choice of RNGs suggests that it is usually not crucial in practice. This assumption has been experimentally verified, \eg, in \citet{rajashekharan2016} for differential evolution and is most certainly due to the fact that modern PRNGs seem to be sufficiently secure and reliable for most practical purposes. The influence of different seeds for a PRNG on various deep learning algorithms for computer vision has been studied empirically in \citet{picard2021} with the result that it is often possible to find seeds that lead to a much better or much worse performance than the average. This highlights the fact that numerical experiments with non-deterministic algorithms have to be conducted carefully to account for the variance of random numbers. However, the specific implications of varying degrees of security and reliability of RNGs on machine learning applications generally remain unresolved, \ie, it generally remains unclear whether a certain machine learning algorithm may suffer or benefit from the artifacts of an imperfect RNG. In the present work, we approach this still rather open field of research by specifically considering the randomness in artificial neural network initialization.

\subsubsection{Quantum RNGs}
As previously stated, quantum computers (or, more generally, quantum systems) have an intrinsic ability to produce truly random outcomes in a way that cannot be predicted or emulated by any classical device \citep{calude2010}. Therefore, it seems natural to utilize them as a source of random numbers in the sense of a quantum random number generator (QRNG). Such QRNGs \citep{herrero2017} have already been realized with different quantum systems, for example using nuclear decay \citep{park2020} or optical devices \citep{nicolo2020}.\par
Summarized, the main difference between randomness from classical systems and randomness from quantum systems is that a classical system is fully deterministic and therefore all randomness can only result from a lack of knowledge about the system, whereas a quantum system is non-deterministic and therefore -- even with perfect knowledge -- an intrinsic randomness may be involved. In this sense, the origin of randomness is different for quantum and classical RNGs. However, it is in principle not possible to mathematically distinguish the randomness of a classical system from the randomness of a quantum system \citep{khrennikov2015}.\par
A simple QRNG can be straightforwardly realized using a quantum circuit. For this purpose, each of its qubits has to be brought into a superposition of 0 and 1 such that both outcomes are equally probable to be measured. This operation can for example be performed by applying a single Hadamard gate on each qubit \citep{nielsen2011}. Each measurement of the circuit consequently generates a sequence of i.i.d. random bits, one for each qubit.\par
However, when computing this simple QRNG circuit on a NISQ device, it can be expected that the results will deviate from the theoretic expectations due to statistical and systematic uncertainties such that the QRNG is likely to produce biased outcomes. This means that it is in fact not guaranteed that the measurement outcomes obey the theoretically predicted probability distribution of a fair coin toss. It is not even guaranteed that the measurement outcomes are truly random in the sense that bits are generated entirely independent. As a consequence (and based on the fact that quantum non-determinism is not ultimately resolved), it cannot be generally taken for granted that random numbers from such a QRNG are naturally ``better'' than random numbers from PRNGs, both with respect to security and reliability. For this reason, technically more refined solutions are necessary to realize trustworthy QRNGs on NISQ decices. Moreover, QRNGs have to be certified similar to classical RNGs. For example, to enable a theoretically secure QRNG, the Bell inequality \citep{pironio2010} or the Kochen-Specker theorem can be utilized \citep{abbott2014,abbott2015,kulikov2017}. For an experimental verification of random bit sequences from a QRNG, entanglement-based public tests of randomness can be used without violating the secrecy of the generated sequences \citep{jacak2020}.\par
Currently, there exist various commercial and non-commercial QRNGs, which can be used to create quantum random numbers on demand, for example \citet{anuqrng2021}. Although there still seem to be some practical challenges \citep{martinez2018,petrov2020}, theoretical and technological advances in the field will most certainly lead to a steady improvement of QRNGs.

\section{Biased QRNG} \label{sec:biased qrng}
Motivated by the work in \citet{bird2020}, we take a different approach than usual in this manuscript. Instead of aiming for a RNG with as little bias as possible, we discuss whether the typical bias in a naively implemented, gate-based QRNG can actually be beneficial for certain machine learning applications. In other words, we consider the bias that is ``naturally'' imposed by the quantum hardware itself (\ie, by the hardware-related errors outlined in \cref{fig:qcp}). In addition to a bias, we also accept that the randomness of the results is not necessarily guaranteed in the sense that the QRNG can (to some degree) produce correlations or predictable patterns from systematic quantum hardware errors. Since the imperfections of the quantum hardware are beyond our control (\ie, they can in particular not be switched off at will), a RNG realized in this way contains unknown and uncontrollable elements. Therefore, we have to analyze its outcomes statistically to capture the effects of these elements on the generated random numbers. In the present section, we first describe our experimental setup for such a naively implemented QRNG and subsequently discuss the statistics of the resulting ``hardware-biased'' quantum random numbers. 

\subsection{Setup}
To realize a hardware-biased QRNG (B-QRNG), we utilize a physical quantum computer, which we access remotely via Qiskit \citep{qiskit2019} using the cloud-based quantum computing service provided by IBM Quantum \citep{ibmq2021}. With this service, users can send online requests for quantum experiments using a high-level quantum circuit model of computation, which are then executed sequentially \citep{larose2019}. The respective quantum hardware, also called \emph{backend}, operates on superconducting transmon qubits.\par
\begin{figure}[!t]
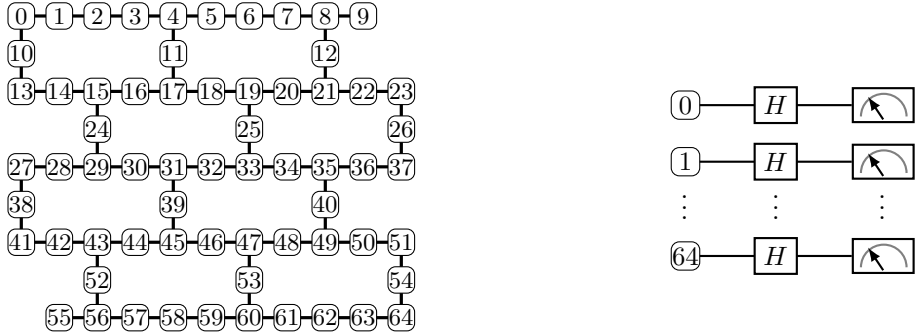

\centering
\begin{subfigure}[t]{.49\linewidth}
\centering
\includestandalone{./fig_backend}
\caption{Topology diagram of the \emph{ibmq\_manhattan} backend with $\num{65}$ qubits. Qubits are shown as boxes with their respective hardware index $n$. Pairs of qubits that support two-qubit gate operations (which are not used in our setup) are connected with lines.}
\label{fig:setup:backend}
\end{subfigure}%
\hfill%
\begin{subfigure}[t]{.49\linewidth}
\centering
\includestandalone{./fig_circuit}
\caption{Circuit diagram. A single Hadamard gate (denoted by $H$) is applied to each of the $\num{65}$ qubits from \subref{fig:setup:backend} and a subsequent measurement is performed. The idealized measurement result of each qubit is either 0 or 1 with an equal probability of 50\%.}
\label{fig:setup:circuit}
\end{subfigure}%
\caption{Main components of our B-QRNG setup: \subref{fig:setup:backend} topology diagram of the backend and \subref{fig:setup:circuit} circuit diagram.}
\label{fig:setup}		
\end{figure}

For our application, we specifically use the \emph{ibmq\_manhattan} backend (version 1.11.1), which is one of the IBM quantum \emph{Hummingbird r2} processors with $N \equiv \num{65}$ qubits. A sketch of the backend topology diagram can be found in \cref{fig:setup:backend}. It indicates the \emph{hardware index} of each qubit and the pairs of qubits that support two-qubit gate operations between them. IBM also provides an estimate for the relaxation time $T_1$ and the dephasing time $T_2$ for each qubit at the time of operation. The mean and standard deviation of these times over all qubits read $T_1 \approx \SI{59.11 +- 15.25}{\micro\second}$ and $T_2 \approx \SI{74.71 +- 31.22}{\micro\second}$, respectively.\par
Initially, all qubits in this backend are prepared in the ground state. Our B-QRNG cicuit, which is sketched in \cref{fig:setup:circuit}, consists of one Hadamard gate applied to each qubit such that it is brought into a balanced superposition of ground state and excited state. A subsequent measurement on each qubit should therefore ideally (\ie, in the error-free case) reveal an outcome of either 0 (corresponding to the ground state) or 1 (corresponding to the excited state) with equal probability. However, since we run the circuit on real quantum hardware, we can expect to obtain random numbers which deviate from these idealized outcomes due to hardware-related errors. An analogous setup with a different backend is considered in \citet{tamura2020,shikano2020,tamura2021}.\par
We sort the qubit measurements according to their respective hardware index in an ascending order so that each run of the backend yields a well-defined bit string of length $N$. Such a single run is called a \emph{shot} in Qiskit. We perform sequences of $S \equiv \num{8192}$ shots (which is the upper limit according to the backend access restrictions imposed by IBM) for which we concatenate the resulting bit strings in the order in which they are executed. Such a sequence of shots is called \emph{experiment} in Qiskit. We repeat this experiment $R \equiv \num{564}$ times ($\num{900}$ experiments is the upper limit set by IBM) and again concatenate the resulting bit strings in the order of execution. A sequence of experiments is denoted as a \emph{job} in Qiskit and can be submitted directly to the backend. It is run in one pass without interruption from other jobs.\par
\begin{figure}[!t]
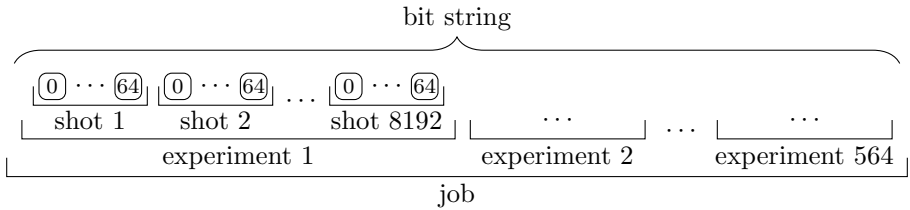

\centering
\includestandalone{./fig_bitstring}
\caption{Bit string composition from our B-QRNG. A single job is submitted to the backend, it consists of $\num{564}$ experiments. In each experiment, $\num{8192}$ shots are performed. In each shot, each of the $\num{65}$ qubits yields a single bit. The resulting bit string consequently contains $\num{300318720}$ bits.}
\label{fig:bitstring}
\end{figure}

Our submitted job ran from March~5, 2021 10:45~AM~GMT to March~5, 2021 11:58~AM~GMT. The final result of the job is a bit string of length $M \equiv NSR=\num{300318720}$ as sketched in \cref{fig:bitstring}. The choice of $R$ is determined by the condition $M \gtrapprox \num{3e8}$, which we have estimated as sufficient for our numerical experiments. We split the bit string into chunks of length $C \equiv \num{32}$ to obtain $L \equiv M/C=\num{9384960}$ random 32-bit integers, which we use for the following machine learning experiments.

\subsection{Statistics}
Before we utilize our generated random numbers for learning algorithms, we first briefly discuss their statistics. The measurement results from the $n$th qubit can be considered as a Bernoulli random variable \citep{forbes2011}, where $n\in\{0,\dots,64\}$ represents the hardware index as outlined in \cref{fig:setup}. Such a variable has a probability mass function
\begin{align} \label{eqn:bernoulli}
f(b;p) \equiv p^b (1-p)^{1-b}
\end{align}
depending on the value of the bit $b \in \mathbb{B}$ and the success probability $p \in [0,1]$ of observing an outcome $b=1$.

\subsubsection{Bias}
We denote the measured bit string from our B-QRNG as a vector $\mathbf{B} \in \mathbb{B}^M$. The extracted bit string exclusively resulting from measurements of the $n$th qubit is given by the vector
\begin{align} \label{eqn:b}
\mathbf{b}_n \equiv \left( B_{n+1}, B_{n+1+N}, \dots, B_{n+1+M-N} \right)
\end{align}
with $\mathbf{b}_n \in \mathbb{B}^{M/N}$. Based on its population, the corresponding expected probability $p_n(0)$ of obtaining the bit $b$ for the $n$th qubit is given by
\begin{align} \label{eqn:pnb}
p_n(b) = \frac{N\sum_{i=1}^{M/N}\mathds{1}_n(i,b)}{M}
\end{align}
with the indicator function
\begin{align}
\mathds{1}_n(i,b) \equiv \begin{cases} 1 & \text{if}\; B_{n+(i-1)N+1} = b \\ 0 & \text{otherwise} \end{cases}
\end{align}
such that $p_n(0) + p_n(1) = 1$. From an idealized prediction of the measurement results of qubits in a balanced superposition, we would assume that all expected probabilities $p_0(b),\dots,p_N(b)$ correspond to the uniform probability 
\begin{align} \label{eqn:pt}
\tilde{p} \equiv \tilde{p}(b) \equiv \frac{1}{2}
\end{align}
with uncertainties coming only from the finite number of samples.\par
\begin{figure}[!t]
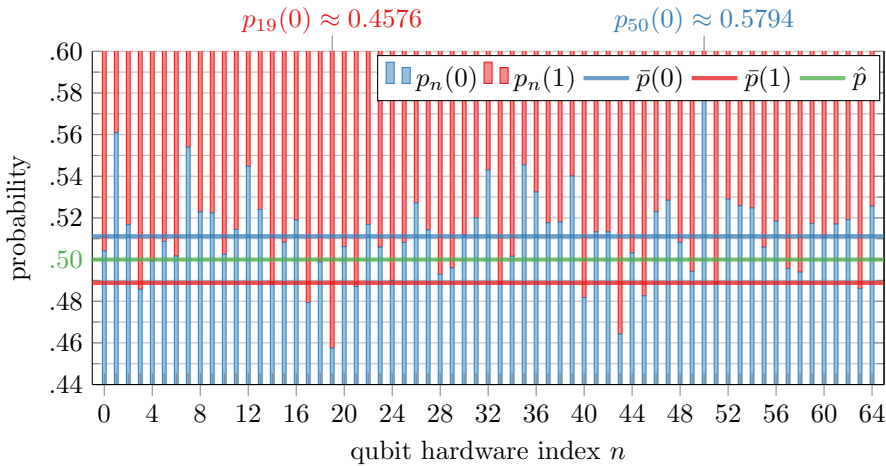

\centering
\includestandalone{./fig_qrng_stats_bits}
\caption{Measured bit distribution for each qubit from the B-QRNG on \emph{ibmq\_manhattan}. We show  the expected probability $p_n(0)$ of obtaining a zero bit from the measured bit string for the $n$th qubit, \cref{eqn:pnb}, and (stacked on top) its complement $p_n(1)=1-p_n(0)$. Also shown are the corresponding expected probabilities with respect to all measured bits $\bar p(0) \approx \num{0.51}$ and $\bar p(1)=1-\bar p(0) \approx \num{0.49}$, respectively, \cref{eqn:pb}. Apparently, all bit distributions deviate differently from the uniform probability $\tilde{p}$, \cref{eqn:pt}, which we assume to be a consequence of the imperfect hardware. The distributions with the highest ($n=50$) and lowest ($n=19$) expected probabilities of obtaining a zero bit are marked on top.}
\label{fig:bitstats}
\end{figure}

We show the estimated probabilities in \cref{fig:bitstats}. It is apparent that all bit probabilities deviate significantly from their idealized value $\tilde{p}$, \cref{eqn:pt}. In particular, we find an expected probability and standard deviation with respect to all measured bits of
\begin{align} \label{eqn:pb}
\bar{p}(0) \approx \num{0.5112 +- 0.0215}.
\end{align}
We assume that this is a consequence of the imperfect hardware with its decoherence and dissipation effects. In particular, the fact that $\bar{p}(0) > \bar{p}(1)$ is most likely a consequence of dissipation since a bit of 0 corresponds to an observation of a qubit ground state, whereas a bit of 1 is associated with an excited state.\par
From a $\chi^2$ test \citep{pearson1900} on the measured bit distribution, the null hypothesis of a uniform zero bit occurrence can be rejected as expected with a confidence level of \num{1.0000}. To further quantify the deviation of the measured probabilities from a uniform distribution, we utilize the discrete Hellinger distance \citep{hellinger1909}
\begin{align} \label{eqn:H}
\mathrm{H}(q_1, q_2) \equiv \frac{1}{\sqrt{2}} \sqrt{ \sum_{i \in Q} \left( \sqrt{q_1(i)} - \sqrt{q_2(i)} \right)^2 },
\end{align}
which can be used to measure similarities between two discrete probability distributions $q_1 \equiv q_1(i)$ and $q_2 \equiv q_2(i)$ defined on the same probability space $Q$. By iterating over all qubits we find the mean and standard deviation
\begin{align}
\langle \mathrm{H}( p_n, \tilde{p} ) \rangle \approx \num{0.0133 +- 0.011}.
\end{align}
The mean value quantifies the average deviation of the measured qubit distributions from the idealized uniform distribution and confirms our qualitative observations. The non-negligible standard deviation results from the fluctuations in-between the individual qubit outcomes.

\subsubsection{Randomness}
Although quantum events intrinsically exhibit a truly random behavior, the output of our B-QRNG is the result of a complex physical experiment behind a technically sophisticated pipeline that appears as a black box to us and it can therefore not be assumed with certainty that its outcomes are indeed statistically independent. To examine this issue in more detail, we briefly study the randomness of the resulting bit string in the following.\par
For this purpose, we make use of the Wald–Wolfowitz runs test \citep{wald1940}, which can be used to test the null hypothesis that elements of a binary sequence are mutually independent. We perform a corresponding test on the measured bit string from the $n$th qubit $\mathbf{b}_n$, \cref{eqn:b}, and denote the resulting p-value as $p^{\mathrm{r}}_n$. The null hypothesis has to be rejected if this probability does not exceed the significance level, which we choose as $\alpha = \num{0.05}$.\par
\begin{figure}[!t]
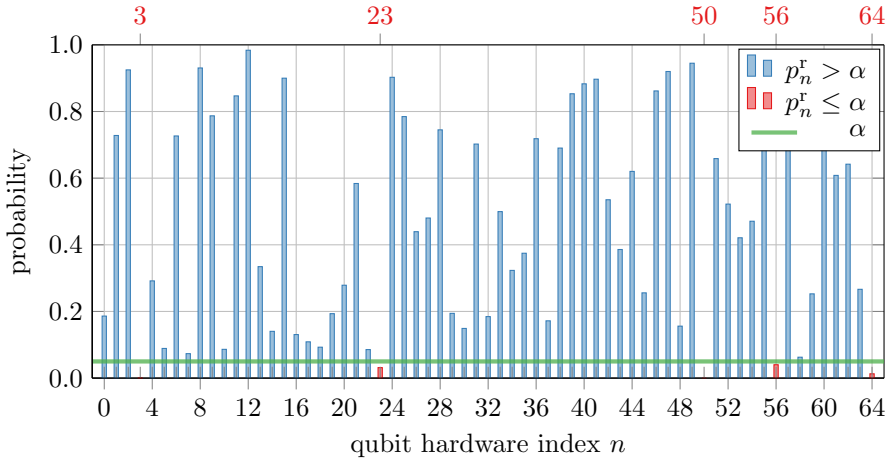

\centering
\includestandalone{./fig_qrng_stats_bits_runs}
\caption{Results of Wald–Wolfowitz runs test on the bit strings of all qubits, where $p^{\mathrm{r}}_n$ denotes the resulting p-value of the bit string of the $n$th qubit $\mathbf{b}_n$, \cref{eqn:b}. We show p-values in different colors depending on whether or not they exceed $\alpha = \num{0.05}$. In case of $p^{\mathrm{r}}_n \leq \alpha$, the corresponding hardware indices are additionally denoted on top of the plot and indicate the qubits that fail the test of randomness.}
\label{fig:bitstatsruns}
\end{figure}

The test results are shown in \cref{fig:bitstatsruns}. We find that the bit strings from almost all qubits pass the test and can therefore be considered random in the sense of the test criteria. However, the bit strings from five qubits fail the test, which implies non-randomness. We also perform a test on the total bit string $\mathbf{B}$, which yields the p-value $p^{\mathrm{r}} \approx \num{0.0000} < \alpha$ such that the test also fails for the entire sequence of random numbers.\par
Summarized, we find that the reliability of the generated quantum random numbers is questionable. A typical binary random sequence from a PRNG of the same length as $\mathbf{B}$ can be expected to pass the Wald–Wolfowitz runs test. However, within the scope of this work, the reason for this observation cannot be further investigated and we accept it as an integral part of our naive approach to the B-QRNG. Further work regarding the properties of our setup (applied to a different quantum hardware) can be found in \citet{tamura2020,shikano2020,tamura2021}, which contain similar observations. A lack of reliability is not surprising considering the fact that we have not aimed for a certified random number generation and our setup is motivated by a strongly idealized model of quantum gate computers, as already mentioned above.

\subsubsection{Integers}
Next, we analyze the resulting random 32-bit integers. To obtain these, we convert $\mathbf{B}$ into a vector of integers $\mathbf{B} \mapsto \mathbf{I} \in \{0,\dots,2^{C}-1\}^L$ by consecutively grouping its elements into bit strings of length $C$ and converting them to non-negative integers according to
\begin{align} \label{eqn:I}
I_j \equiv \sum_{i=0}^{C-1} B_{C (j-1) + i + 1} 2^i
\end{align}
with $j \in \{1,\dots,L\}$. For a bit string of Bernoulli random variables $\mathbf{B}$ with a fair success probability $p=\tilde{p}$, \cref{eqn:bernoulli,eqn:pt}, the sequence of random integers in $\mathbf{I}$ would be uniformly distributed. However, as we have seen before, this assumption does not hold true for the results from our B-QRNG. So the question arises as to what the distribution of random integers looks like for our unfair set of Bernoulli variables.\par
For this purpose, we rescale the elements of $\mathbf{I}$ by a division by 
\begin{align} \label{eqn:xi}
	\xi \equiv 2^{C}-1
\end{align}
such that $\mathbf{I}/\xi \in [0,1]^L$ and group the range $[0,1]$ into $K \equiv \num{250}$ equally sized bins. Thus, the population of the $k$th bin is given by
\begin{align} \label{eqn:cm}
c_k \equiv \sum_{i=1}^{L} \mathds{1}(I_i,k)
\end{align}
with the indicator function
\begin{align} \label{eqn:cm:1}
\mathds{1}(i,k) \equiv \begin{cases} 1 & \text{if}\; k < K \;\land\; \frac{k-1}{K} \leq \frac{i}{\xi} < \frac{k}{K} \\ 1 & \text{if}\; k = K \;\land\; \frac{K-1}{K} \leq \frac{i}{\xi} \\ 0 & \text{otherwise} \end{cases}
\end{align}
for $k \in \{1,\dots,K\}$.\par
Additionally, we consider a simplified theoretical description of the bin population by modeling the bit string as the result of a Bernoulli process with a single success probability $p$, \cref{eqn:bernoulli}. That is, the bits represent \iid Bernoulli random variables. The integer $j \in \{0,\dots,\xi\}$ corresponding to a bit string $\boldsymbol{\tau}(j) \in \mathbb{B}^{C}$ is determined in analogy to \cref{eqn:I} such that $\sum_{i=0}^{C-1} \tau_{i+1}(j) 2^i = j$. The probability mass function of the resulting integers can consequently be written as
\begin{align} \label{eqn:I:P}
P(j,p) \equiv \prod_{i=1}^{C} p^{\tau_i(j)} (1-p)^{1-{\tau_i(j)}}.
\end{align}
Its expected value is given by
\begin{align} \label{eqn:I:e}
\hat{I}(p) \equiv \sum_{i=0}^{\xi} i P(i,p) = \xi p
\end{align}
and the information entropy \citep{shannon1948} in \emph{nats} by 
\begin{align} \label{eqn:I:S}
S_I(p) \equiv \sum_{i=0}^{\xi} P(i,p) \ln P(i,p) = - C \left[ p \ln p + (1-p) \ln (1-p) \right].
\end{align}
We show a plot of \cref{eqn:I:e,eqn:I:S} in \cref{fig:simintstatssummary}. Finally, the predicted (possibly non-integer) population of the $k$th bin reads
\begin{align} \label{eqn:ct}
\hat{c}_k(p) \equiv L \sum_{i=0}^{\xi} \mathds{1}(i,k) P(i,p),
\end{align}
which we use as our simplified model of \cref{eqn:cm}.\par
\begin{figure}[!t]
\centering
\includestandalone{./fig_sim_stats_ints_summary}
\caption{Expected value $\hat{I}(p)$, \cref{eqn:I:e}, and entropy $S_I(p)$, \cref{eqn:I:S}, for a random integer from the domain $\{0,\dots,\xi\}$ resulting from a string of $C$ random bits from a Bernoulli process with success probability $p$, \cref{eqn:bernoulli}. The expected value is proportional to $p$, whereas the entropy attains its maximum value at $p=\num{0.5}$. We apply rescaling factors to constrain both quantities to the same scale.}
\label{fig:simintstatssummary}
\end{figure}
\begin{figure}[!t]
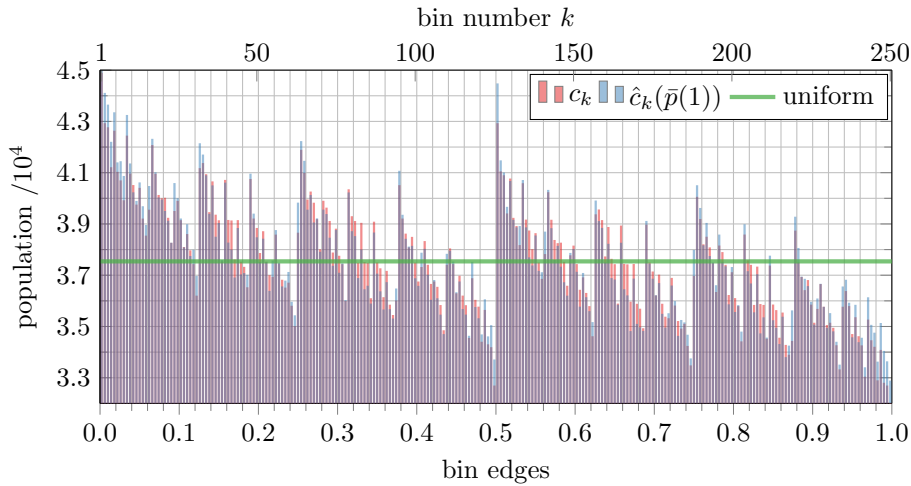

\centering
\includestandalone{./fig_qrng_stats_ints}
\caption{Measured distribution of 32-bit integers from the B-QRNG. The values from the generated vector of random integers $\mathbf{I}$, \cref{eqn:I}, are rescaled by a division by $(2^{32}-1)$ and sorted into $\num{250}$ equally sized bins. The $k$th bin (with $k \in \{1,\dots,250\}$) has a population of $c_k$ according to \cref{eqn:cm}. For comparison, the corresponding theoretic bin population of the $k$th bin $\hat{c}_k(\bar{p}(b))$ is shown, which is obtained from a Bernoulli process according to \cref{eqn:ct} with a success probability of $p=\bar{p}(1)=1-\bar{p}(0)$, \cref{eqn:pb}. The minor deviations between the two populations results from the finite number of measured samples as well as the observation that bits from different qubits have their own success probability, \cf \cref{fig:bitstats}. An outline of the uniform bin population is shown as a frame of reference.}
\label{fig:intstats}
\end{figure}

We show both the measured bin population $c_k$, \cref{eqn:cm}, and the theoretical bin population $\hat{c}_k(p)$, \cref{eqn:ct}, for a success probability $p$ corresponding to the expected probability of all measured bits $\bar{p}(1)=1-\bar p(0)$, \cref{eqn:pb}, in \cref{fig:intstats}. Clearly, the generated sequence of random integers is not uniformly distributed (\ie, with a population of $L/K$ in each bin). Instead, we find a complex arrangement of spikes and valleys in the bin populations.\par
Specifically, since $\bar{p}(0) > \bar{p}(1)$, random integers become more probable when their binary representation contains as many zeros as possible, which is reflected in the bin populations. In particular, the first bin (containing the smallest integers) has the highest population. The minor deviations between the measured and the theoretic bin populations results from the finite number of measured samples and the simplification of the theoretical model: The success probability of each bit from the B-QRNG specifically depends on the qubit it is generated from as shown in \cref{fig:bitstats}, whereas our theoretical model only uses one success probability for all bits corresponding to $\bar{p}(1)$.\par
We recall the Hellinger distance, \cref{eqn:H}, to quantify the deviation of the distribution of integers from the uniform distribution. Specifically, we find
\begin{align} \label{eqn:H:int}
\mathrm{H}(p_c, \tilde{p}_c) \approx \num{0.0213},
\end{align}
where we have made use of the measured integer distribution $p_c \equiv p_c(k) \equiv c_k/L$ and the corresponding uniform distribution $\tilde{p}_c \equiv \tilde{p}_c(k) \equiv 1/K$ with $k\in\{1,\dots,K\}$. This metric quantifies our observations from \cref{fig:intstats}.\par
For comparative purposes, we show additional theoretical bin populations for other success probabilities in \cref{fig:simintstats}. As expected, the rugged pattern of the distribution becomes sharper for lower or higher values of $p$ and the deviation from the uniform distribution increases.
\begin{figure}[!t]
\centering
\includestandalone{./fig_sim_stats_ints}
\caption{Theoretical distribution of 32-bit integers in analogy to \cref{fig:bitstats} for different success probabilities $p \in \{p_1=\num{0.3},p_2=\num{0.4},p_3=\num{0.5},p_4=\num{0.6},p_5=\num{0.7}\}$, \cref{eqn:ct}. The population axis is scaled logarithmically. We also show the (rescaled) mean values $\hat{I}(p)/\xi$, \cref{eqn:I:e}, and the uniform distribution $\tilde{p}_c$ as used in \cref{eqn:H:int}. The corresponding Hellinger distances, \cref{eqn:H}, with $\hat{p}_c(p) \equiv \hat{p}_c(p;k) \equiv \hat{c}_k(p)/L$ and $k\in\{1,\dots,K\}$ read $\mathrm{H}(\hat{p}_c(p_1), \tilde{p}_c) \approx \mathrm{H}(\hat{p}_c(p_5), \tilde{p}_c) \approx \num{0.3776}$, $\mathrm{H}(\hat{p}_c(p_2), \tilde{p}_c) \approx \mathrm{H}(\hat{p}_c(p_4), \tilde{p}_c) \approx \num{0.1867}$, and $\mathrm{H}(\hat{p}_c(p_3), \tilde{p}_c) \approx \num{0.0000}$, respectively.}
\label{fig:simintstats}
\end{figure}

\section{Experiments} \label{sec:experiments}
To study the effects of quantum-based network initializations, we consider two independent experiments, which are both implemented in \emph{PyTorch} \citep{pytorch2019}: First, a convolutional neural network (CNN) and second, a recurrent neural network (RNN). The choice of these experiments is motivated by the statement from \citet{bird2020} that ``neural network experiments show greatly differing patterns in learning patterns and their overall results when using PRNG and QRNG methods to generate the initial weights.''\par
To ensure repeatability of our experiments, PyTorch is run in \emph{deterministic mode} with fixed (\ie, hard-coded) random seeds. The main hardware component is a \emph{Nvidia GeForce GTX 1080 Ti} graphics card. Our Python implementation of the experiments is publicly available online \citep{implementation2021}.\par
In the present section, we first summarize the considered RNGs. Subsequently, we present the two experiments and discuss their results.

\subsection{RNGs} \label{sec:experiments:rngs}
\begin{figure}[!t]
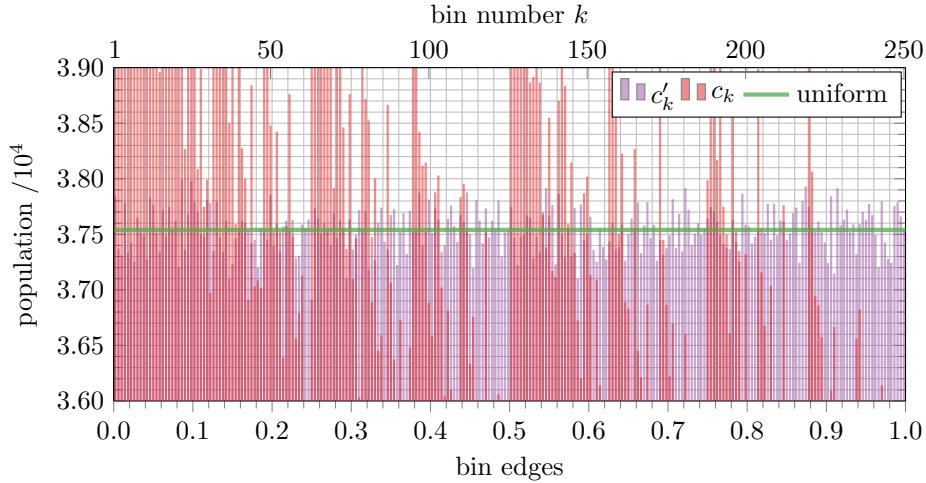

\centering
\includestandalone{./fig_anu_stats_ints}
\caption{Distribution of 32-bit integers from the QRNG in analogy to \cref{fig:intstats}. The values from the vector of random integers $\mathbf{I'}$ are rescaled by a division by $(2^{32}-1)$ and sorted into $\num{250}$ equally sized bins. The population of the $k$th bin (with $k \in \{1,\dots,250\}$) is denoted by $c'_k$. For comparison, we also show the corresponding population $c_k$, \cref{eqn:cm}, from the B-QRNG and an outline of the uniform bin population.}
\label{fig:anuintstats}
\end{figure}

In total, we use four different RNGs to initialize neural network weights:
\begin{enumerate}
\item B-QRNG: Our hardware-biased quantum random number generator introduced in \cref{sec:biased qrng} from which we extract the integer sequence $\mathbf{I}$ according to \cref{eqn:I}. The data is publicly available online \citep{bqrng2023}.
\item QRNG: A bias-free quantum random number generator \citep{anuqrng2021} based on quantum-optical hardware that performs broadband measurements of the vacuum field contained in the radio-frequency sidebands of a single-mode laser to produce a continuous stream of binary random numbers \citep{symul2011,haw2015}. We particularly use a publicly available pre-generated sequence of random bits from this stream \citep{anudata2017}, extract the first $M$ bits and convert them into the integer sequence $\mathbf{I'} \in \{0,\dots,2^{C}-1\}^L$ according to \cref{eqn:I}. Based on the Hellinger distance $\mathrm{H}(p'_c, \tilde{p}_c) \approx \num{0.0018}$, \cref{eqn:H}, with $p'_c \equiv p'_c(k) \equiv c'_k/L$ and $c'_k \equiv \sum_{i=1}^{L} \mathds{1}(I'_i,k)$, \cref{eqn:cm:1}, for $k\in\{1,\dots,K\}$, we find that $\mathbf{I'}$ is indeed much closer to the uniform distribution than $\mathbf{I}$, \cref{eqn:H:int}. We visualize the corresponding integer distribution in \cref{fig:anuintstats}.
\item PRNG: The (presumably unbiased) native pseudo-random number generator from PyTorch.
\item B-PRNG: A ``pseudo hardware-biased quantum random number generator'', which generates a bit string of \iid Bernoulli random variables with a success probability $p$ corresponding to the expected probability of all measured bits $\bar{p}(1)=1-\bar{p}(0)$, \cref{eqn:bernoulli,eqn:pb}, using the native pseudo-random number generator from PyTorch. The bit strings are then converted into integers according to \cref{eqn:I}. Their probability mass function is given by \cref{eqn:I:P}.
\end{enumerate}
All of these RNGs, which are summarized in \cref{tab:rngs}, produce 32-bit random numbers. However, the random numbers from the B-QRNG and the QRNG are taken in order (\ie, unshuffled) from the predefined sequences $\mathbf{I}$ and $\mathbf{I'}$, respectively, whereas the PRNG and the B-PRNG algorithmically generate random numbers on demand based on a given random seed.\par
\begin{table}[!t]
\centering
\caption{Overview over the four considered RNGs presented in \cref{sec:experiments:rngs}, which are either based on a classical pseudo-random number generator or a quantum experiment (as indicated by the rows) and yield either unbiased or biased outcomes (as indicated by the columns).}
\label{tab:rngs}	
\setlength\extrarowheight{1pt}
\begin{tabular}{>{\centering\arraybackslash}p{1.6cm}>{\centering\arraybackslash}p{2cm}>{\centering\arraybackslash}p{1.75cm}}
\toprule
& unbiased & biased \\
\midrule
classical & PRNG & B-PRNG \\
quantum   & QRNG & B-QRNG \\
\bottomrule
\end{tabular}
\end{table}
For the sake of completeness, we also analyze the binary random numbers from the B-QRNG and the QRNG, respectively, with the NIST Statistical Test Suite for the validation of random number generators \citep{rukhin2010,nist2010}. For this purpose, the bit strings are segmented into smaller sequences and multiple statistical tests are evaluated on each sequence. Each test consists of one or more sub-tests with the null hypothesis that the sequence being tested is random. Based on the proportion of sequences for which a sub-test satisfies the null hypothesis, it is considered as passed or rejected, where a rejection indicates non-randomness. A more detailed discussion about this procedure can also be found in \citet{sys2015}.\par
A summary of our results is listed in \cref{tab:nist}. It shows that the B-QRNG numbers fail a majority of statistical tests of randomness, as expected, whereas the QRNG passes all.
\begin{table}[!pt]
\centering
\caption{Summary of the results from the NIST Statistical Test Suite for the validation of random number generators \citep{nist2010} applied to the whole sequence of binary random numbers from the B-QRNG and the QRNG, respectively. For each of the two bit strings, a series of statistical tests is evaluated, where each test consists of one or more sub-tests with the null hypothesis that the sequence being tested is random. The bit strings are segmented into sequences and each sub-test is run on each of these sequences. A sub-test is accepted or rejected based on a certain proportion of sequences that satisfy the null hypothesis. A rejection therefore indicates non-randomness. We choose \num{100} bitstreams in the program options such that each sequence contains at least $10^6$ bits, as recommended. The required number of passed sequences of the total number of sequences for the acceptance of a sub-test is \num{96} of \num{100} for all tests except for the Random excursion tests, for which it is \num{68} of \num{72} due to a reduced number of effectively used sequences. The predefined standard parameters are used for all tests (\eg, significance level $\alpha = \num{0.01}$). We list for each statistical test the corresponding number of accepted (``\accept'') and rejected (``\reject'') sub-tests. The total number of acceptances and rejections are shown at the bottom in bold. In addition, the column ``\passed'' contains information about the number of passed sequences. For tests with only one or two sub-tests, we explicitly list the number of passed sequences. Otherwise, we present the respective means and standard deviations. A detailed description of the software and its statistical tests of randomness can be found in \citet{rukhin2010}. Summarized, the QRNG data passes all tests of randomness, whereas most of the tests fail for the B-QRNG data, which indicates non-randomness. Note that the Random excursions tests are not applicable for the B-QRNG data since they require the acceptance of the rejected Frequency test \citep{sys2015}.
}
\label{tab:nist}	
\setlength\extrarowheight{1pt}
\begin{tabularx}{\textwidth}{Xcccccc}
\toprule
\multicolumn{1}{l}{Test name} & {\accept} & {\reject} & {\passed} & {\accept} & {\reject} & {\passed} \\
\midrule
%
{Approximate entropy                  } & {\num{  0}} & {\num{  1}} & { \num{0} } & {\num{  1}} & {\num{  0}} & { \num{99} } \\\hline
{Frequency within block               } & {\num{  0}} & {\num{  1}} & { \num{0} } & {\num{  1}} & {\num{  0}} & { \num{98} } \\\hline
{Cumulative sums                      } & {\num{  0}} & {\num{  2}} & { \num{0}, \num{0} } & {\num{  2}} & {\num{  0}} & { \num{98}, \num{99} } \\\hline
{Discrete Fourier transform           } & {\num{  0}} & {\num{  1}} & { \num{60} } & {\num{  1}} & {\num{  0}} & { \num{100} } \\\hline
{Frequency                            } & {\num{  0}} & {\num{  1}} & { \num{0} } & {\num{  1}} & {\num{  0}} & { \num{99} } \\\hline
{Linear complexity                    } & {\num{  1}} & {\num{  0}} & { \num{100} } & {\num{  1}} & {\num{  0}} & { \num{96} } \\\hline
{Longest run of ones within block     } & {\num{  0}} & {\num{  1}} & { \num{72} } & {\num{  1}} & {\num{  0}} & { \num{100} } \\\hline
{Non-overlapping template matching    } & {\num{ 40}} & {\num{108}} & { \num{52.4} $\pm$ \num{41.5} } & {\num{148}} & {\num{  0}} & { \num{99.0} $\pm$ \num{1.0} } \\\hline
{Overlapping template matching        } & {\num{  0}} & {\num{  1}} & { \num{0} } & {\num{  1}} & {\num{  0}} & { \num{100} } \\\hline
{Random excursions                    } & {\num{  0}} & {\num{  0}} & --- & {\num{  8}} & {\num{  0}} & { \num{71.5} $\pm$ \num{0.7} } \\\hline
{Random excursions variant            } & {\num{  0}} & {\num{  0}} & --- & {\num{ 18}} & {\num{  0}} & { \num{71.7} $\pm$ \num{0.6} } \\\hline
{Binary matrix rank                   } & {\num{  1}} & {\num{  0}} & { \num{100} } & {\num{  1}} & {\num{  0}} & { \num{100} } \\\hline
{Runs                                 } & {\num{  0}} & {\num{  1}} & { \num{0} } & {\num{  1}} & {\num{  0}} & { \num{96} } \\\hline
{Serial                               } & {\num{  1}} & {\num{  1}} & { \num{0}, \num{100} } & {\num{  2}} & {\num{  0}} & { \num{98}, \num{99} } \\\hline
{Maurer's ``universal statistical''   } & {\num{  0}} & {\num{  1}} & { \num{80} } & {\num{  1}} & {\num{  0}} & { \num{99} } \\\hline
{\textbf{{Total}}} & {\textbf{\num{ 43}}} & {\textbf{\num{119}}} & & {\textbf{\num{188}}} & {\textbf{\num{  0}}} & \\
\midrule
{} & \multicolumn{2}{c}{B-QRNG} & & \multicolumn{2}{c}{QRNG} & \\
\bottomrule
\end{tabularx}
\end{table} 

\subsection{CNN} \label{sec:experiments:cnn}
In the first experiment, we consider a \emph{LeNet-5} inspired CNN with ReLU activation functions and without dropout \citep{lecun1998a}. The network weights are initialized as proposed by \citet{he2015}, but we use a uniform distribution instead of a normal distribution, as is also common. This means that each weight $w_i$ (with $i=1,2,\dots$) is sampled uniformly according to
\begin{align} \label{eqn:wh}
	w_i \sim [-h_i,h_i],
\end{align}
where $h_i>0$ is chosen such that a constant output variance can be achieved over all layers. The network biases are initialized analogously.\par
As data we use the MNIST handwritten digit recognition problem \citep{lecun1998b}, which contains \num{70000} grayscale images of handwritten digits in $\num{28}\times\num{28}$ pixel format. The digits are split into a training set of \num{60000} images and a training set of \num{10000} images. The network is trained using \emph{Adadelta} \citep{zeiler2012} over $d \equiv \num{14}$ epochs.\par
\begin{figure}[!t]
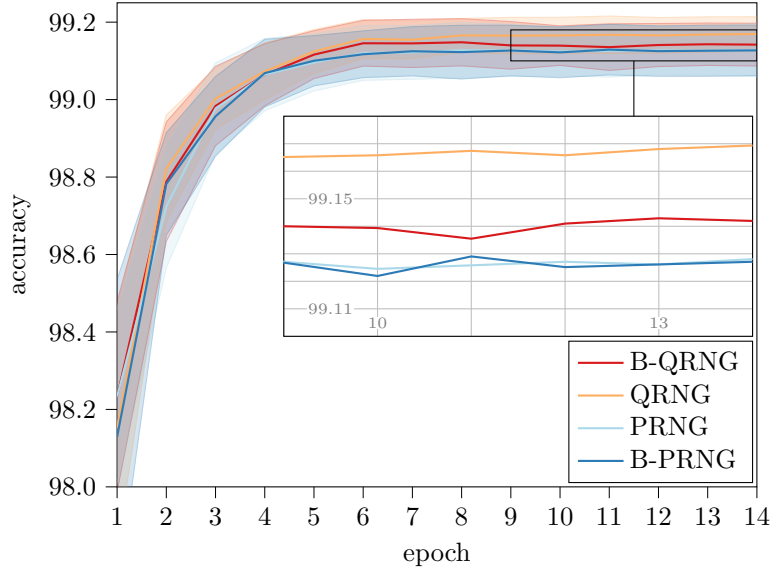

\centering
\includestandalone{./fig_cnn_eval}
\caption{CNN test accuracy convergence on the MNIST data set using four different random number generators (B-QRNG, QRGN, PRGN and B-PRNG from \cref{sec:experiments:rngs}). Shown are mean values over \num{31} runs with the respective standard deviations (one sigma). The inset plot zooms in on the means of the final epochs.}
\label{fig:CNNLeNetMNIST}
\end{figure}

In \cref{fig:CNNLeNetMNIST} we show the CNN test accuracy convergence for each epoch over \num{31} independent training runs using the four RNGs from \cref{sec:experiments:rngs}. The use of a biased RNG means that the \citeauthor{he2015} initialization is actually effectively realized based on a non-uniform distribution instead of a uniform distribution. Therefore, such an approach could potentially be considered a new type of initialization strategy (depending on the bias), which is why one might expect a different training efficiency. However, the results show that the choice of RNG for the network weight initialization has no major effect on the CNN test accuracy convergence. Only a closer look reveals that the mean QRNG results seem to be slightly superior to the others in the last epochs.\par
To quantify this observation, we utilize Welch's (unequal variances) $t$-test for the null hypothesis that two independent samples have identical expected values without the assumption of equal population variance \citep{welch1947}. We apply this test to two of each of the four results from different RNGs, where the resulting test accuracies from all runs in a specific epoch are treated as samples. We denote the two results to be compared as $\mathbf{x}$ and $\mathbf{y}$, respectively, with $\mathbf{x},\mathbf{y} \in \mathbb{R}^{\num{31} \times d}$ for \num{31} runs and $d$ epochs. Consequently, for each pair of results and each epoch $i \in \{1,\dots,d\}$, we obtain a two-tailed p-value $p_i^t(\mathbf{x}, \mathbf{y})$. The null hypothesis has to be rejected if such a p-value does not exceed the significance level, which we choose as $\alpha = \num{0.05}$.\par
We are particularly interested whether the aforementioned hypothesis holds true for all epochs. To counteract the problem of multiple comparisons, we use the Holm-Bonferroni method \citep{holm1979} to adjust the p-values $p_i^t(\mathbf{x}, \mathbf{y}) \mapsto \bar{p}_i^t(\mathbf{x}, \mathbf{y})$ for all $i \in \{1,\dots,d\}$. Summarized, if the condition
\begin{align} \label{eqn:ttest}
	\min_{i, \mathbf{x}, \mathbf{y}} \bar{p}_i^t(\mathbf{x}, \mathbf{y}) \equiv \min_{\mathbf{x}, \mathbf{y}} \bar{p}_{\min}^t(\mathbf{x}, \mathbf{y}) \overset{!}{>} \alpha = \num{0.05}
\end{align}
is fulfilled, no overall statistically significant deviation between the results from different RNGs is present.\par
In addition, we also quantify the correlation of $\mathbf{x}$ and $\mathbf{y}$ using the Pearson correlation coefficient \citep{pearson1895}
\begin{align} \label{eqn:rho}
	\rho(\mathbf{x},\mathbf{y}) \equiv \frac{\sum_{i=1}^c \bar{x}^m_i \bar{y}^m_i}{\sqrt{\sum_i (\bar{x}^m_i)^2 \sum_j (\bar{y}^m_j)^2}} \in [-1,1]
\end{align}
of the mean values over all runs, where we make use of the abbreviations $\bar{x}'_i \equiv x'_i - \sum_{i=1}^d x'_i/d$, $x'_i \equiv \sum_{j=1}^{\num{31}} x_{ji}/\num{31}$, $\bar{y}'_i \equiv y'_i - \sum_{i=1}^d y'_i/d$, and $y'_i \equiv \sum_{j=1}^{\num{31}} y_{ji}/\num{31}$. A coefficient of \num{1} implies a perfect linear correlation of the means, whereas a coefficient of \num{0} indicates no linear correlation.\par
For the results from the CNN experiment, we obtain the similarity and correlation metrics listed in \cref{tab:experimentstats} in the rows marked with ``CNN''. Summarized, we find a high mutual similarity (\cref{eqn:ttest} holds true) and almost perfect mutual correlations of the results. This means that the choice of RNG for the network weight initialization has no statistically significant effect on the CNN test accuracy convergence and, in particular, the QRNG results are not superior despite the visual appearance in \cref{fig:CNNLeNetMNIST}. 
\begin{table}[!t]
\centering
\caption{Minimum p-values from Welch's $t$-test over all epochs $\bar{p}_{\min}^t(\mathbf{x}, \mathbf{y})$, \cref{eqn:ttest}, and Pearson correlation coefficient $\rho(\mathbf{x}, \mathbf{y})$, \cref{eqn:rho}, of the experimental data. The metrics are listed for all mutual combinations of the results from the four RNGs (B-QRNG, QRGN, PRGN, and B-PRNG from \cref{sec:experiments:rngs}) of all experiments (CNN, RNN-M, and RNN-A from \cref{sec:experiments:cnn,sec:experiments:rnn}, respectively).}
\label{tab:experimentstats}	
\setlength\extrarowheight{1pt}
\begin{tabular}{>{\centering\arraybackslash}p{0.5cm}>{\centering\arraybackslash}p{1.5cm}>{\centering\arraybackslash}p{1.5cm}>{\centering\arraybackslash}p{1.5cm}>{\centering\arraybackslash}p{1.5cm}}
\toprule
& $\mathbf{x}$ & $\mathbf{y}$ & $\bar{p}_{\min}^t(\mathbf{x}, \mathbf{y})$ & $\rho(\mathbf{x}, \mathbf{y})$  \\\noalign{\vskip\doublerulesep\vskip-\arrayrulewidth}
%
\midrule
{\multirow{6}{*}{\cellrot{\rotatebox[origin=c]{90}{CNN}}}} & B-QRNG & QRGN & \num{0.3784} & \num{0.9984} \\
{} & B-QRNG  & PRGN & \num{1.0000} & \num{0.9980} \\
{} & B-QRNG  & B-PRNG & \num{0.9749} & \num{0.9986} \\
{} & QRGN & PRGN & \num{0.0641} & \num{0.9941} \\
{} & QRGN & B-PRNG & \num{0.0577} & \num{0.9992} \\
{} & PRGN & B-PRNG & \num{1.0000} & \num{0.9951} \\
\midrule
{\multirow{6}{*}{\cellrot{\rotatebox[origin=c]{90}{RNN-M}}}} & B-QRNG & QRGN & \num{1.0000} & \num{0.9997} \\
{} & B-QRNG  & PRGN & \num{0.4526} & \num{0.9995} \\
{} & B-QRNG  & B-PRNG & \num{1.0000} & \num{0.9998} \\
{} & QRGN & PRGN & \num{1.0000} & \num{0.9998} \\
{} & QRGN & B-PRNG & \num{1.0000} & \num{0.9999} \\
{} & PRGN & B-PRNG & \num{0.5355} & \num{0.9996} \\
\midrule
{\multirow{6}{*}{\cellrot{\rotatebox[origin=c]{90}{RNN-A}}}} & B-QRNG & QRGN & \num{1.0000} & \num{0.9946} \\
{} & B-QRNG  & PRGN & \num{1.0000} & \num{0.9954} \\
{} & B-QRNG  & B-PRNG & \num{1.0000} & \num{0.9962} \\
{} & QRGN & PRGN & \num{1.0000} & \num{0.9944} \\
{} & QRGN & B-PRNG & \num{1.0000} & \num{0.9951} \\
{} & PRGN & B-PRNG & \num{1.0000} & \num{0.9965} \\
\bottomrule
\end{tabular}
\end{table} 	
At this point, the question arises whether a different bias of the RNGs might have led to better training results. To answer this question, we consider additional pseudo-random number generators B-PRNG($p$), which are based on a Bernoulli process with success probability $p$, \cref{eqn:bernoulli}, such that the originally considered B-PRNG corresponds to B-PRNG($\bar{p}(1)$), \cref{eqn:pb}, and the PRNG corresponds to B-PRNG($\num{0.5}$). In the extreme cases of $p=\num{0}$ and $p=\num{1}$, B-PRNG($p$) is not random anymore and produces only constant values of $0$ and $2^{32}-1$, respectively. The probability mass function of the resulting integers is given by \cref{eqn:I:P}. We train the CNN again on the MNIST data set with a weight initialization based on B-PRNG($p$) for different values of $p \in [0,1]$ and consider the test accuracy at epoch \num{14}.\par
The results are shown in \cref{fig:biasCNN}. Clearly, the mean test accuracy attains a maximum at $p=\num{0.5}$, which corresponds to an unbiased pseudo-random number generator (\ie, the PRGN). For smaller and larger success probabilities, the mean test accuracy decreases. In particular, we observe a steep drop in performance for $p<\num{0.2}$ and $p>\num{0.95}$, which indicates that a bias of the random number generator towards 0 has more severe effects than a bias towards 1. The worst performance is achieved for $p=\num{0}$ and $p=\num{1}$, respectively.\par
We recall that for $p=\num{0.5}$, weights are sampled uniformly around zero, \cref{eqn:wh}. Thus, for $p>\num{0.5}$, the weights are more probable to be positive, whereas for $p<\num{0.5}$, they are more probable to be negative, \cf \cref{eqn:I:e}. Since our CNN contains ReLU activation functions, a shift of the weights towards negative values leads to vanishing gradients. According to our experiments, this seems to become significant for $p<\num{0.2}$. On the other hand, an equivalent shift towards positive values does not drastically decrease the training performance and even for $p=\num{0.95}$ the test accuracy is above $\SI{98.6}{\percent}$. However, for $p=\num{1}$ the test accuracy also drops. We think that the reason for this behavior is that the weights are in this case constant and attain the maximum value of the distribution, \cref{eqn:wh}. The resulting lack of diversity, which is for example evident from the entropy, \cref{eqn:I:S}, is probably the cause for the bad training performance \citep{frankle2019}.
\begin{figure}[!t]
\centering
\begin{subfigure}[t]{.49\linewidth}
\centering
\includestandalone{./fig_bias_cnn_1}
\caption{full bias range $p \in [0,1]$}
\label{fig:biasCNN:1}
\end{subfigure}%
\hfill	
\begin{subfigure}[t]{.49\linewidth}
\centering
\includestandalone{./fig_bias_cnn_2}
\caption{zoom on $p \in [\num{0.2},\num{0.95}]$}
\label{fig:biasCNN:2}
\end{subfigure}%
\caption{CNN test accuracy on the MNIST data at epoch \num{14} using different pseudo-random number generators B-PRNG($p$), which are based on a Bernoulli process with success probability $p$, \cref{eqn:bernoulli}. We consider $p \in \{\num{0}, \num{.05}, \num{.1}, \num{.2}, \num{.3}, \num{.4}, \num{.5}, \num{.6}, \num{.7}, \num{.8}, \num{.9}, \num{.95}, \num{1}\}$. Shown are mean values over \num{30} runs with the respective standard deviations (one sigma) as error bars for \subref{fig:biasCNN:1} the full bias range and \subref{fig:biasCNN:2} a zoom on the peak of the accuracy at $p=\num{0.5}$. For comparison, we also plot the corresponding results from \cref{fig:CNNLeNetMNIST} for the B-PRNG with $p=\bar{p}(1)$, \cref{eqn:pb}, as well as for the PRNG with $p=\num{0.5}$.}
\label{fig:biasCNN}		
\end{figure}

\subsection{RNN} \label{sec:experiments:rnn}
In the second experiment, we consider a recurrent LSTM cell with a uniform initialization in analogy to \cref{eqn:wh}, which we apply on the synthetic adding and memory standard benchmarks \citep{hochreiter1997} with $T=\num{64}$ for the memory problem. For this purpose, we use \emph{RMSprop} \citep{hinton2012} with a step size of \num{e-3} to optimize LSTM cells \citep{hochreiter1997} with a state size of \num{256}. For each problem, a total of \num{9e5} updates with training batches of size \num{128} is computed until the training stops. In total, there are $\lfloor \num{9e5} / \num{128} \rfloor = \num{7031}$ training steps.\par
Since the synthetic data sets are infinitely large, overfitting is not an issue and we can consequently use the training loss as performance metric. Specifically, we consider \num{89} consecutive training steps as one epoch, which leads to $d \equiv \num{4687}/\num{89} = \num{79}$ epochs in total, each associated with the mean loss of the corresponding training steps.\par
\begin{figure}[!t]
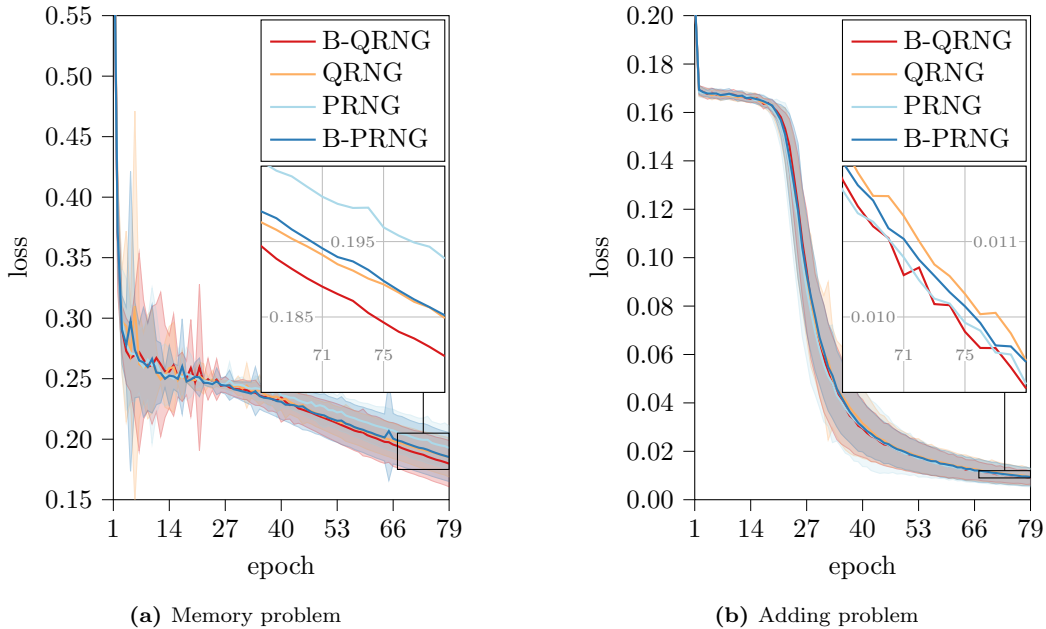

\centering
\begin{subfigure}[t]{.49\linewidth}
\centering
\includestandalone{./fig_rnn_eval_memory}
\caption{Memory problem}
\label{fig:RNN:mem}
\end{subfigure}%
\hfill	
\begin{subfigure}[t]{.49\linewidth}
\centering
\includestandalone{./fig_rnn_eval_adding}
\caption{Adding problem}
\label{fig:RNN:add}
\end{subfigure}%
\caption{RNN convergence on two benchmark data sets using four different RNGs (B-QRNG, QRGN, PRGN and B-PRNG from \cref{sec:experiments:rngs}). Shown are mean values over \num{31} runs with the respective standard deviations (one sigma) in analogy to \cref{fig:CNNLeNetMNIST}. The inset plot zooms in on the means of the final epochs.}
\label{fig:RNN}		
\end{figure}

The results are shown in \cref{fig:RNN}, where we present the loss for each of the \num{79} epochs over \num{31} independent training runs for both problems. Again, we compare the results using random numbers from the four RNGs from \cref{sec:experiments:rngs}. The use of a biased RNG effectively realizes a non-uniform initialization (depending on the bias) in comparison with the uniform initialization from a non-biased RNG. However, we find that no RNG yields a major difference in performance.\par
In analogy to the first experiment, we list the similarity and correlation metrics in \cref{tab:experimentstats} in the rows marked with ``RNN-M'' and ``RNN-A'', respectively. Again, we find a high mutual similarity (\cref{eqn:ttest} holds true) and correlation. Thus, the choice of RNG also has no statistically significant effect in this second experiment. Due to the numerical effort required to train the RNNs, we cannot perform an analysis of different biases of RNGs as in the first experiment.

\section{Conclusions} \label{sec:conclusions}
Summarized, by running a naively designed quantum random number generator on a quantum gate computer, we have generated a random bit string. Its statistical analysis has revealed a significant bias and mutual dependencies as imposed by the quantum hardware. When converted into a sequence of integers, we have found a specially shaped distribution of values with a rich pattern. We have utilized these integers as hardware-biased quantum random numbers (B-QRNG). Motivated by the results from \citet{bird2020}, we have deliberately chosen to use these biased and correlated random numbers to study their impact on machine learning algorithms.\par
Specifically, we have studied their effect on the initialization of artificial neural network weights in two experiments. For comparison, we have additionally considered unbiased random numbers from another quantum random number generator (QRNG) and a classical pseudo-random number generator (PRNG) as well as random numbers from a classical pseudo-random number generator replicating the hardware bias (B-PRNG). The two experiments consider a CNN and a RNN, respectively, and show no statistically significant influence of the choice of RNG.\par
Despite a similar setup, we have not been able to replicate the observation from \citet{bird2020}, where it is stated that quantum random number generators and pseudo-random number generators ``do inexplicably produce different results to one another when employed in machine learning.'' However, we have not explicitly attempted to replicate the numerical experiments from the aforementioned work, but have instead considered two different examples that we consider typical applications of neural networks in machine learning.\par
Since our results are only exemplary, it may indeed be possible that there is an advantage in the usage of biased quantum random numbers for certain applications. Based on our studies, we expect, however, that in such cases it will in fact not be the ``true randomness'' of the quantum random numbers, but rather the opposite -- their hardware-induced bias, including possible correlations -- that will cause an effect. But is quantum hardware really necessary to produce such results? It seems that classical pseudo-random number generators are also able to mimic these effects. Even more, because the reliability and security of PRNGs can be ensured with less effort and a greater confidence than that of gate-based QRNGs on NISQ devices. Therefore, we think that for typical machine learning applications the usage of (high-quality) pseudo-random numbers is sufficient. Accordingly, a more elaborate experimental or theoretical study of the effects of biased pseudo-random numbers (with particular patterns) on certain machine learning applications could be a suitable research topic, \eg, to better understand the claims from \citet{bird2020}.\par
Repeatability is generally difficult to achieve for numerical calculations involving random numbers \citep{crane2018}. In particular, our B-QRNG can in principle not be forced to reproduce a specific random sequence (as opposed to PRNGs). Furthermore, the statistics of the generated quantum random numbers may depend on the specific configuration of the quantum hardware at the time of operation. It might therefore be possible that a repetition of the numerical experiments with quantum random numbers obtained at a different time or from a different quantum hardware may lead to significantly different results. To ensure the greatest possible transparency, the source code for our experiments is publicly available online \citep{implementation2021} and may serve as a point of origin for further studies.

\begin{acknowledgements}
We thank Christian Bauckhage and Bogdan Georgiev for informative discussion. Parts of this work have been funded by the Federal Ministry of Education and Research of Germany as part of the competence center for machine learning ML2R (01$\vert$S18038A and 01$\vert$S18038B), the Fraunhofer Cluster of Excellence Cognitive Internet Technologies (CCIT), the Fraunhofer Research Center Machine Learning as well as the State of North Rhine-Westphalia (Germany) as part of the Lamarr Institute for Machine Learning and Artificial Intelligence. We thank the University of Bonn for access to its \emph{Auersberg} and \emph{Teufelskapelle} clusters and acknowledge the use of IBM Quantum services for this work. Access to the IBM hardware has been funded by the Ministry of Science and Health of the State of Rhineland-Palatinate (Germany) as part of the project AnQuC-2. For our numerical calculations, we have made particular use of \citet{scipy2020,statsmodels2010,pytorch2019}.
\end{acknowledgements}

\makeatletter\@ifclassloaded{svjour3}{\input{suppl_declaration}}{}\makeatother 

\reftoc\bibliography{main}

\end{document}